# Observational Evidence of Magnetic Reconnection in the Terrestrial Foreshock Region

K. Jiang[1], S. Y. Huang[1*], H. S. Fu[2], Z. G. Yuan[1], X. H. Deng[3], Z. Wang[2], Z. Z. Guo[2], S. B. Xu[1], Y. Y. Wei[1], J. Zhang[1], Z. H. Zhang[1], Q. Y. Xiong[1], and L. Yu[1]

[1]School of Electronic Information, Wuhan University, Wuhan, China

[2]School of Space and Environment, Beihang University, Beijing, China

[3]Institute of Space Science and Technology, Nanchang University, Nanchang, China

[*]Corresponding author: shiyonghuang@whu.edu.cn

**Abstract:**

Electron heating/acceleration in the foreshock, by which electrons may be energized beyond thermal energies prior to encountering the bow shock, is very important for the bow shock dynamics. And then these electrons would be more easily injected into a process like diffusive shock acceleration. Many mechanisms have been proposed to explain electrons heating/acceleration in the foreshock. Magnetic reconnection is one possible candidate. Taking advantage of the Magnetospheric Multiscale mission, we present two magnetic reconnection events in the dawn-side and dusk-side ion foreshock region, respectively. Super-Alfvénic electron outflow, demagnetization of the electrons and the ions, and crescent electron distributions in the plane perpendicular to the magnetic field are observed in the sub-ion-scale current sheets. Moreover, strong energy conversion from the fields to the plasmas and significant electron temperature enhancement are observed. Our observations provide direct evidence that magnetic reconnection could occur in the foreshock region and heat/accelerate the electrons therein.

## 1. Introduction

Magnetic reconnection, as one of the most common and most important physical phenomena in astrophysical and laboratory plasmas, can efficiently convert magnetic energy to the particles' energy and change the topology of the magnetic field simultaneously. It has been observed in different environments like the Earth's magnetotail (e.g., Øieroset et al., 2001; Borg et al., 2005; Fu et al., 2017; Huang et al., 2010, 2012a, 2012b, 2016a, 2016b, 2018; Jiang et al., 2019; Torbert et al., 2018; Zhou et al., 2019), magnetopause (e.g., Sonnerup et al., 1981; Deng & Matsumoto 2001; Burch et al., 2016; Zhou et al., 2017; Wang et al., 2019a; Fu et al., 2019), magnetosheath (e.g., Retinò et al., 2007; Phan et al., 2018), transition region of the bow shock (e.g., Chen et al., 2019; Gingell et al., 2019; Wang et al., 2019b), the sheath of interplanetary coronal mass ejection (e.g., Huang et al., 2012c). The foreshock, located at the upstream region of the bow shock and extending to a dozen $R_E$ ($R_E$ is the Earth's radius), is a wide region with a turbulent and nonlinear magnetic field. Particles with energy up to hundreds of keV have been observed in the foreshock (e.g., Wilson et al., 2016). How the particles are heated/accelerated in the foreshock is a longstanding question. Many mechanisms have been proposed to explain it: Fermi acceleration (Liu et al., 2017; Turner et al., 2018), Betatron acceleration (Liu et al., 2019a), lower hybrid drift instability (Zhang et al., 2010), etc. Magnetic reconnection is also one possible mechanism. Matsumoto et al. (2015) performed a particle-in-cell (PIC) simulation and found that suprathermal electrons can be accelerated by magnetic islands that are generated by magnetic reconnection in quasi-perpendicular shock. Wang et al. (2019b) and Gingell et al. (2019) observed magnetic reconnection events occurring in the transition region of the bow shock. Meanwhile, electron heating was also observed. Their observations suggest that magnetic reconnection may contribute to shock energy dissipation (Wang et al., 2019b; Gingell et al., 2019). Chen et al. (2019) have reanalyzed the previous magnetic reconnection events in the transition region of the bow shock using the First Order Taylor Expansion (FOTE) method (Fu et al., 2015, 2016), and confirmed the existence of magnetic nulls in these magnetic reconnection events. Bai et al. (2020) observed an ion-scale flux rope at the trailing edge of a hot flow anomaly, as one type of foreshock transients, in the upstream region of the quasi-parallel bow shock. They inferred that the flux rope might be generated by magnetic reconnection locally. And they suggested that magnetic reconnection and flux ropes might play a role in particle acceleration/heating inside the foreshock transients (Bai et al., 2020). Furtherly, Liu et al. (2020) showed two magnetic reconnection events in the foreshock transients and Wang et al. (2020) observed reconnection current sheets in short large-amplitude magnetic structures.

In this work, we present two electron-only reconnection events in the dawn-side and dusk-side ion foreshock region, respectively. These two events did not take place in the foreshock transients like the work mentioned before but in the usual foreshock. The features of the observed magnetic reconnection events in the foreshock are similar to those in other environments like the magnetosphere and magnetosheath (e.g., Phan et al., 2018; Lu et al., 2020; Man et al., 2020; Zhou et al., 2021; Zhong et al., 2021). Importantly, electron heating/acceleration is observed. These observations imply that magnetic reconnection can occur in the foreshock like in other astrophysical environments, and the heated/accelerated electrons may further affect the bow shock dynamics.

## 2. MMS Observations

The data used in this work are provided by the instruments onboard the Magnetospheric Multiscale (MMS) mission. The magnetic field recorded by Fluxgate Magnetometer (FGM, Russell et al., 2016), the electric field measured by Electric Double Probe (EDP, Ergun et al., 2016; Lindqvist et al., 2016), and 3D particle velocity distributions and the plasma moments collected by the Fast Plasma Instrument (FPI, Pollock et al., 2016) are used in the present study.

Figure 1 is a sketch of the dayside near Earth's space environment (Balogh & Treumann 2013). The magnetic field in the foreshock is turbulent. The magenta rectangles in the dawn-side and the dusk-side foreshock denote two magnetic reconnection events observed by MMS at around 10:10:02 UT on 2016 December 8 (case 1) and 08:38:39 UT on 2017 January 26 (case 2), respectively.

Figure 2 exhibits the overview for both cases with data in burst mode in geocentric solar ecliptic (GSE) coordinates. One can see that MMS crossed the bow shock from the magnetosheath to the foreshock for case 1 and from the foreshock to the magnetosheath for case 2. The foreshock regions are characterized by backstreaming ions (e.g., Wilson et al., 2014) from the bow shock (Figure 2a and 2i), and lower density compared to the magnetosheath (Figure 2f and 2n). This implies that MMS were located in the foreshock region (marked by magenta bars on the top of Figure 2). The magnetosheath regions are characterized by a decrease of velocity in the x direction with respect to the foreshock, heated electrons (Figure 2h and 2p), higher density (Figure 2f and 2n), and larger fluctuations in the magnetic field (Figure 2c and 2k) compared to the foreshock. The transition regions between the foreshock and the magnetosheath are identified as the bow shock (e.g., Wang et al., 2019b). Two vertical dashed lines mark the reconnection events in the foreshock. Taking the interval from 10:02 to 10:04 UT (from 09:49 to 09:50 UT) in case 1 and from 08:18 to 08:19 UT (from 08:42 to 08:43 UT) of case 2 as the upstream (downstream) of the bow shock, the angles between the upstream magnetic field and the bow shock normal ($\theta_{Bn}$), calculated with a mixed method (Abraham-Shrauner 1972; Schwartz 1998), are 19° for case 1 and 33° for case 2, implying that both case 1 and case 2 were observed in the upstream of the quasi-parallel bow shock. In the foreshock region, there are many types of foreshock transients, for example: hot flow anomalies (HFAs) (e.g., Schwartz et al., 1985; Zhang et al., 2010), foreshock bubbles (FBs) (e.g., Omidi et al., 2010; Turner et al., 2013), etc. These foreshock transients have hot and tenuous cores with the sharp increase in the solar wind velocity and are bound by the compressional boundaries. The plasma temperature, density, and velocity didn't exhibit such typical characteristics of the foreshock transients in both case 1 and case 2. Thus, we can conclude that both reconnection events occur in the usual foreshock, not in the foreshock transients.

Figure 3 displays the details of case 1 in the dusk-side foreshock on 2016 December 8 when MMS were located at [11.3, 3.7, 0.5] $R_E$ in GSE coordinates. The maximum distance among MMS is 7.5 km during this time interval. The data recorded by all MMS are nearly the same, thus we only present the data from MMS4 here. Minimum variance analysis (MVA, Sonnerup and Scheiblev 1998) is performed on the electron velocity from 10:10:02.46 to 10:10:02.67 UT to establish the local boundary normal (LMN) coordinates. The three eigenvectors are $L$=[-0.34, 0.71, 0.62], $M$=[0.77, 0.59, -0.25], and $N$=[-0.55, 0.39, -0.74], where $N$ is the normal direction of the current sheet. The ratio of the maximum eigenvalue and the median eigenvalue is 5, the ratio of the median eigenvalue and the minimum eigenvalue is 344. Timing analysis (Russell et al., 1983) is performed on $B_y$=0 (from 10:10:02.59 to 10:10:02.61 UT). The result from timing analysis (287×[-0.72, 0.20, -0.66] km s$^{-1}$ in GSE coordinates) has an angle of 15.3° with respect to the normal direction determined by MVA, implying that the results are consistent with each other. All vectors are presented in LMN coordinate in Figure 3.

One can see a current sheet from 10:10:02.4 to 10:10:02.8 UT with intense current density (up to 872 nA m$^{-2}$ in Figure 3f). The width of the current sheet estimated from multiplying the duration of the bipolar change of $B_l$ (from 10:10:02.48 to 10:10:02.63 UT) and the current sheet velocity from timing analysis is ~43 km (i.e., ~0.7 $d_i$ ~28.7 $d_e$, where $d_i$=64.7 km is ion inertial length and $d_e$=1.5 km is electron inertial length calculated with average upstream plasma parameters $N_i$=12.4 cm$^{-3}$ from 10:08:30 to

10:09:30 UT), indicating that this is a sub-ion-scale current sheet. There are bipolar electron jets in L direction around the center of the current sheet (the maximum is 771 km s$^{-1}$ and the minimum is -192 km s$^{-1}$). When the background ion flows are subtracted, the velocities of electron jets are 451 km s$^{-1}$ and -469 km s$^{-1}$ in L direction (Figure 3d), which are much larger than ion Alfvén speed $V_a$ (~83 km s$^{-1}$ derived from average upstream plasma parameters |$B$|=13.4 nT and $N_i$=12.4 cm$^{-3}$ from 10:08:30 to 10:09:30 UT). Thus, they are identified as super-Alfvénic electron jets. $V_{em}$ is strong with a peak up to 352 km s$^{-1}$, which contributes to a strong unipolar out-of-plane $J_m$ (Figure 3f). However, the ion flow doesn't change significantly (Figure 3c). With the bipolar change of $V_{el}$, $B_l$ changes from negative to positive then return to negative, $B_m$ has a positive-negative-positive change (Figure 3a), which is consistent with the Hall magnetic field if we consider the trajectory shown in Figure 3o. The negative $B_m$ near the $B_l$ reversal could be an excursion into the inflow region. The number density of electrons increases asymmetrically in the current sheet (Figure 3b). In the foreshock, the velocity of the background plasma is significant, which leads to a result that the electric field is dominated by the convection component. To investigate the electric field caused by the local process, the convection component should be subtracted. The electric field in the ion frame is shown in Figure 3g. The reconnection electric field ($\boldsymbol{E}$ + $\boldsymbol{V}_i \times \boldsymbol{B}$)$_m$ nearly stays positive during the crossing the current sheet. There is a clear bipolar change in ($\boldsymbol{E}$ + $\boldsymbol{V}_i \times \boldsymbol{B}$)$_n$, which is consistent with the Hall electric field. ($\boldsymbol{E}$ + $\boldsymbol{V_i} \times \boldsymbol{B}$)$_\mathbf{n}$ doesn't show a tripolar change with $B_l$, which may because that MMS have been passed the diffusion region when $B_l$ crosses zero for the second time. Distinct deviations of the M components of the perpendicular ion and electron velocity from the M component of the $\boldsymbol{E} \times \boldsymbol{B}$ drift motion (Figure 3h) denote the demagnetization of the ions and electrons. The energy conversion ($\boldsymbol{J} \cdot \boldsymbol{E}'$, where $\boldsymbol{E}' = \boldsymbol{E} + \boldsymbol{V}_e \times \boldsymbol{B}$, Figure 3i) at the center of the current sheet is positive with a maximum of 2.2 nW m$^{-3}$, implying that energy is converted from the fields to the plasmas. We also examined the electron distribution functions during the crossing of the current sheet (Figure 3j-3l). Obvious crescent distributions are found in the plane perpendicular to the ambient magnetic field. All clues prove that MMS encountered a magnetic reconnection event in the dusk-side ion foreshock. Moreover, no obvious ion outflow was observed, implying that MMS encountered an electron-only reconnection event (Phan et al., 2018). The trajectory of MMS4 crossing the reconnection event is inferred and shown in Figure 3o.

We utilize the FOTE method to examine whether there are magnetic nulls in case 1. FOTE method is used to calculate the distances between magnetic nulls and MMS ($R$) and identify the types of magnetic nulls with and without a guide field (Fu et al., 2015). By means of diagonalization, one can obtain the three eigenvalues of the Jacobian matrix $\delta \boldsymbol{B}$. Different combinations of eigenvalues stand for different types of magnetic nulls. Magnetic nulls are divided into two types: radial nulls (A, B, and X) and spiral nulls (As, Bs, and O, e.g., Fu et al., 2015). A, B, As and Bs are 3D nulls. They can degenerate into 2D structures (X and O) when one of the eigenvalues is zero. The results of the FOTE method can be trusted only when the following conditions are satisfied: 1) $R_{min}$ (min{$\boldsymbol{R}_{MMS1}$ - $\boldsymbol{R}_{null}$, $\boldsymbol{R}_{MMS2}$ - $\boldsymbol{R}_{null}$, $\boldsymbol{R}_{MMS3}$ - $\boldsymbol{R}_{null}$, $\boldsymbol{R}_{MMS4}$ - $\boldsymbol{R}_{null}$}) is less than 1 $d_i$ (ion inertial length); 2) $\eta=|\nabla \cdot \boldsymbol{B}|/|\nabla \times \boldsymbol{B}|$ and $\xi=|\nabla \cdot \boldsymbol{B}|/|\lambda|_{max}$ are less than 40%. Details about the FOTE method can be found in Fu et al. (2015, 2016) and Liu et al. (2019b). The results derived from FOTE method are presented in Figure 3m and 3n. The $R_{min}$ is less than 1 $d_i$ (horizontal black dashed line in Figure 3m), meanwhile, $\eta$ and $\xi$ are much smaller than 40% after 10:10:02.6 UT, which implies that the results are reliable. Besides, the finding of As and Bs nulls in the current sheet indicates that there are null pairs (e.g., Pontin, 2011). Such null pairs in the reconnection diffusion region have also been observed by Deng et al., (2009) and Fu et al., (2019). The magnetic null's topologies can be reconstructed by FOTE method (Fu et al., 2015, 2016, 2017). The 3D magnetic topology (Figure 3p) and 2D view (Figure 3q) along ($e_1$, 0, 0) of case 1 (reconstructed at 10:10:02.607 UT, when the points are at the |$\boldsymbol{B}$|

minimum and $\eta$ and $\zeta$ are smallest nearby, as marked by the cyan arrow in Figure 3n) are presented in the eigenvector coordinates for better revealing the topological features of the magnetic nulls. The eigenvector coordinates ($e_1$, $e_2$, $e_3$) are obtained from the Jacobian matrix $\delta \boldsymbol{B}$. In the eigenvector coordinates, a few points of the magnetic field around the nulls are traced to obtain the topology. The reconstructed topologies show one X-line consisting of a spine and a fan plane during the observations, consistent with the theoretical models (e.g., Lau & Finn 1990), which further supports that MMS detected a reconnection event.

Moreover, electron temperature $T_e$ (both $T_{e\|}$ and $T_{e\perp}$) increases during the crossing of the current sheet (Figure 3e), especially in $T_{e\|}$ (maximum of $T_{e\|}$ increase 23% compared to the background plasma averaged from 10:10:02.2 to 10:10:02.4 UT), indicating clear electron heating in the reconnecting current sheet. These observations provide direct evidence that electron-only reconnection can occur in the foreshock region, and can heat/accelerate electrons therein.

The second case was observed by MMS in the dawn-side foreshock around 08:38:39 UT on 2017 January 26 when MMS were located at [10.9, -3.5, 1.8] $R_E$ in GSE coordinates. The maximum separation between MMS is 6.1 km during this time interval. Only MMS1's data are presented here due to the nearly the same data recorded by all MMS spacecraft. LMN coordinates are established by MVA on the electron velocity (from 08:38:39.49 to 08:38:39.67 UT). The eigenvectors are $\boldsymbol{L}$=[-0.42, 0.75, 0.52], $\boldsymbol{M}$=[-0.03, -0.58, 0.81] and $\boldsymbol{N}$=[0.91, 0.32, 0.26] in GSE coordinates. The ratio of the maximum eigenvalue and the median eigenvalue is 15, and the ratio of the median eigenvalue and the minimum eigenvalue is 77, implying the reliability of the results. The results of timing analysis of the magnetic field (performed at $B_z$=0, from 08:38:39.61 to 08:38:39.64 UT) are 198×[-0.85, -0.37, -0.38] km s$^{-1}$ (GSE), which is the velocity along the normal direction to the current sheet.

The angle of the normal direction determined by timing analysis and $N$ from MVA is 8.2°, indicating that the results from the two methods are consistent.

Figure 4 displays the details of case 2 in LMN coordinates. A current sheet with a peak up to -942 nA m$^{-2}$ is observed from 08:38:39.3 to 08:38:39.9 UT (Figure 4f). The width of the current sheet is 27.7 km (~0.4 $d_i$, ~18.5 $d_e$, here $d_i$=63.7 km is ion inertial length and $d_e$=1.5 km is electron inertial length calculated with average plasma density $N_i$=12.8 cm$^{-3}$ from 08:36 to 08:38 UT), implying that the current sheet is a sub-ion-scale structure. There are bipolar electron jets around the center of the current sheet (up to 290 km s$^{-1}$ and -299 km s$^{-1}$, in which the background flows have been subtracted, Figure 4d), which are super-Alfvénic electron jets ($V_a$=66.5 km s$^{-1}$ is ion Alfvén speed calculated with the average plasma parameters $|B|$=10.9 nT and $N_i$=12.8 cm$^{-3}$ from 08:36 to 08:38 UT). A strong out-of-plane velocity ($V_{em}$, up to 277 km s$^{-1}$) is observed when $V_{el}$ bipolar change (Figure 4d). Around 08:38:39.6 UT, $B_l$ changes from positive to negative then returns to positive, $B_m$ has a negative-0-negative (relative to a guide field ~5 nT) change (Figure 4a), which is consistent with the Hall magnetic field. The number density of electrons asymmetrically increases in the current sheet with a dip around the center of the current sheet (Figure 4b). The reconnection electric field ($E + V_i \times B$)$_m$, supposed to be unipolar, changes its sign after ~08:38:39.6 UT (Figure 4g), which may be polluted by ($E + V_i \times B$)$_n$ due to the inaccuracy of the LMN coordinates (Burch et al., 2020) or due to the temporal or spatial effect from the out-of-plane electric field $E_m$ (Dong et al., 2021). The electric field without the convection component ($E + V_i \times B$, Figure 4g) shows a clear bipolar change in ($E + V_i \times B$)$_n$, which is consistent with the Hall electric field. The unambiguous discrepancy between the M components of the ion and electron perpendicular velocity and the M component of the $E \times B$ drift motion (Figure 4h) indicates that ions and electrons are demagnetized. Strong energy conversion from the fields to the plasmas is also observed at the center of the current sheet with a maximum value up to 5.9 nW m$^{-3}$ (Figure 4i). Obvious crescent distributions in the plane perpendicular to the ambient magnetic field are observed (Figure 4j-4l). In addition, there is no clear change in ion flow during the current sheet crossing (Figure 4c), implying the absence of ion outflow therein. Therefore, MMS detected an electron-only reconnection event (Phan et al., 2018) in the dawn-side foreshock. The inferred trajectory of MMS1 crossing the reconnection event is shown in Figure 4o. The FOTE method is also used to examine and reconstruct the topology of the magnetic nulls in this case. Radial nulls in the current sheet (Figure 4m and 4n) and clear reconstructed X-line topology (Figure 4p and 4q, reconstructed at 08:38:39.628 UT, when the points are at the |$\boldsymbol{B}$| minimum and $\eta$ and $\zeta$ are smallest nearby, as marked by the cyan arrow in Figure 4n) are observed, which further supports that MMS detected a reconnection event. An unambiguous increase of electron temperature (both parallel and perpendicular temperature) is also observed, more significantly in electron parallel temperature (maximum of $T_{e\parallel}$ increase 32.4% compared to the background plasma averaged from 08:38:39 to 08:38:39.3 UT, Figure 4e).

## 3. Conclusions

With high spatiotemporal resolution data from the MMS mission, two electron-only reconnection events in the usual foreshock region are identified in this study. Super-Alfvénic electron outflow, decoupling of electrons and ions from the magnetic field, and crescent electron distributions in the plane perpendicular to the ambient magnetic field are observed during the crossings of sub-ion-scale reconnecting current sheets. Strong energy conversion from the fields to the plasmas and clear electron temperature increase are detected, which indicates that the electrons can be heated/accelerated by magnetic reconnection in the foreshock. The FOTE method detects reliable X points during the crossing of the current sheets, providing further evidence for magnetic reconnection in the foreshock region.

In the foreshock, magnetic reconnection and its products (like flux rope) have only been observed in the foreshock transients (Bai et al., 2020; Liu et al., 2020), which requires a preexistent discontinuity in the solar wind. However, it has never been observed in the usual foreshock region. Thus, our observations imply that magnetic reconnection is universally occurring in the usual foreshock region, and maybe a universal process for electron heating/acceleration in the foreshock. More observations and further investigations with the aid of kinetic simulations should be performed to investigate magnetic reconnection in the foreshock region and its influence on the foreshock dynamics.

**Acknowledgements**

This work was supported by the National Natural Science Foundation of China (41874191, 42074196, 41925018) and the National Youth Talent Support Program. We thank the entire MMS team and instrument leads for data access and support. MMS data are publicly available from the MMS Science Data Center at https://lasp.colorado.edu/mms/sdc/public/about/browse-wrapper/.

**Figure captions**

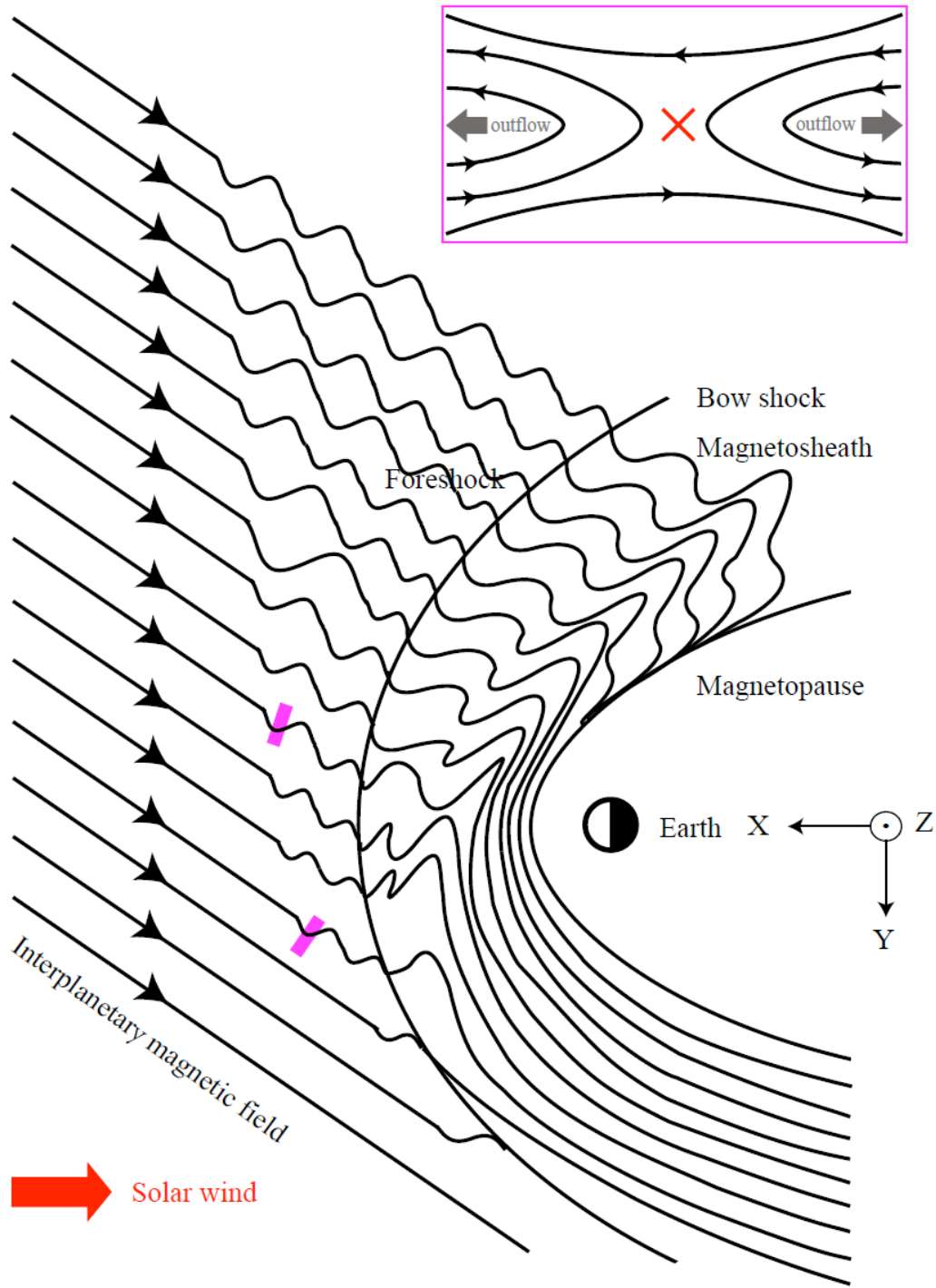

**Figure 1.** Sketch of dayside Earth's space environment. The black lines with arrows represent the magnetic field and the magenta rectangles represent the magnetic reconnection events observed by MMS with a zoom-in electron-only reconnection pattern in the top right corner.

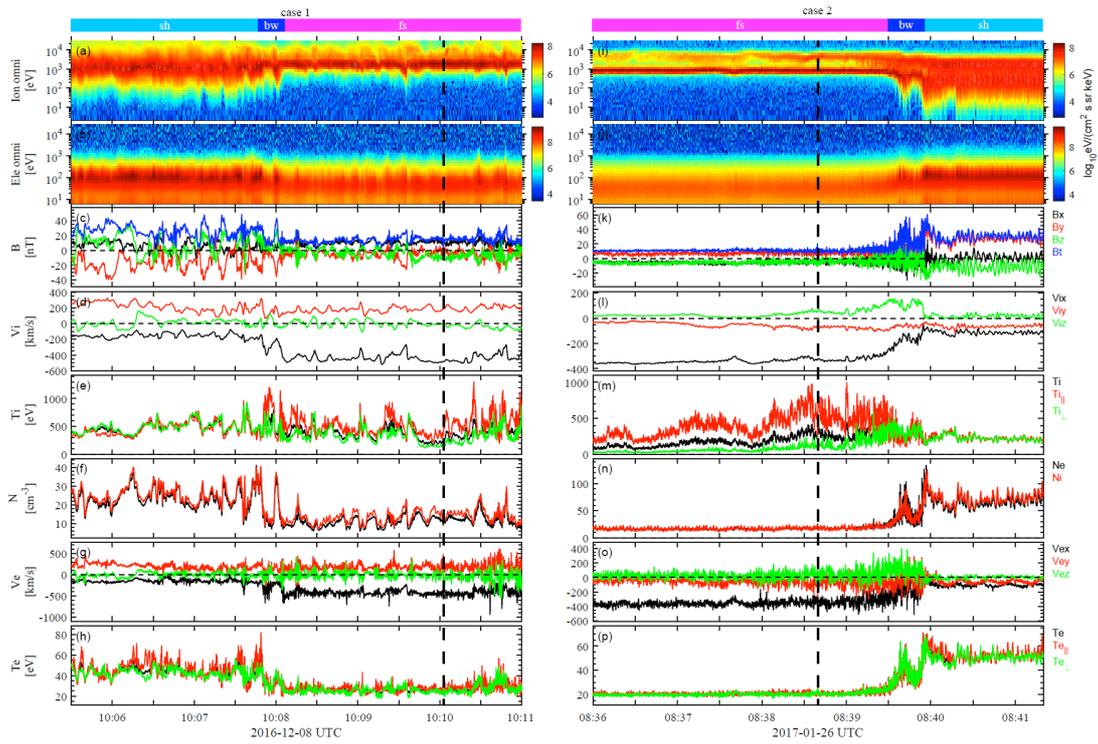

**Figure 2.** Overview of the environments of two magnetic reconnection events in the foreshock region with data in burst mode. (a and i) Ion omni differential energy flux, (b and j) electron omni differential energy flux, (c and k) magnetic field, (d and l) $V_i$, (e and m) $T_i$, (f and n) $N_e$ and $N_i$, (g and o) $V_e$, (h and p) $T_e$. All data are from MMS1 and presented in GSE coordinates. The bars on the top mark the different regions, in which "sh" stands for the magnetosheath, "bw" stands for the bow shock, and "fs" stands for the foreshock. The vertical black dashed lines mark the time intervals when case 1 and case 2 were analyzed.

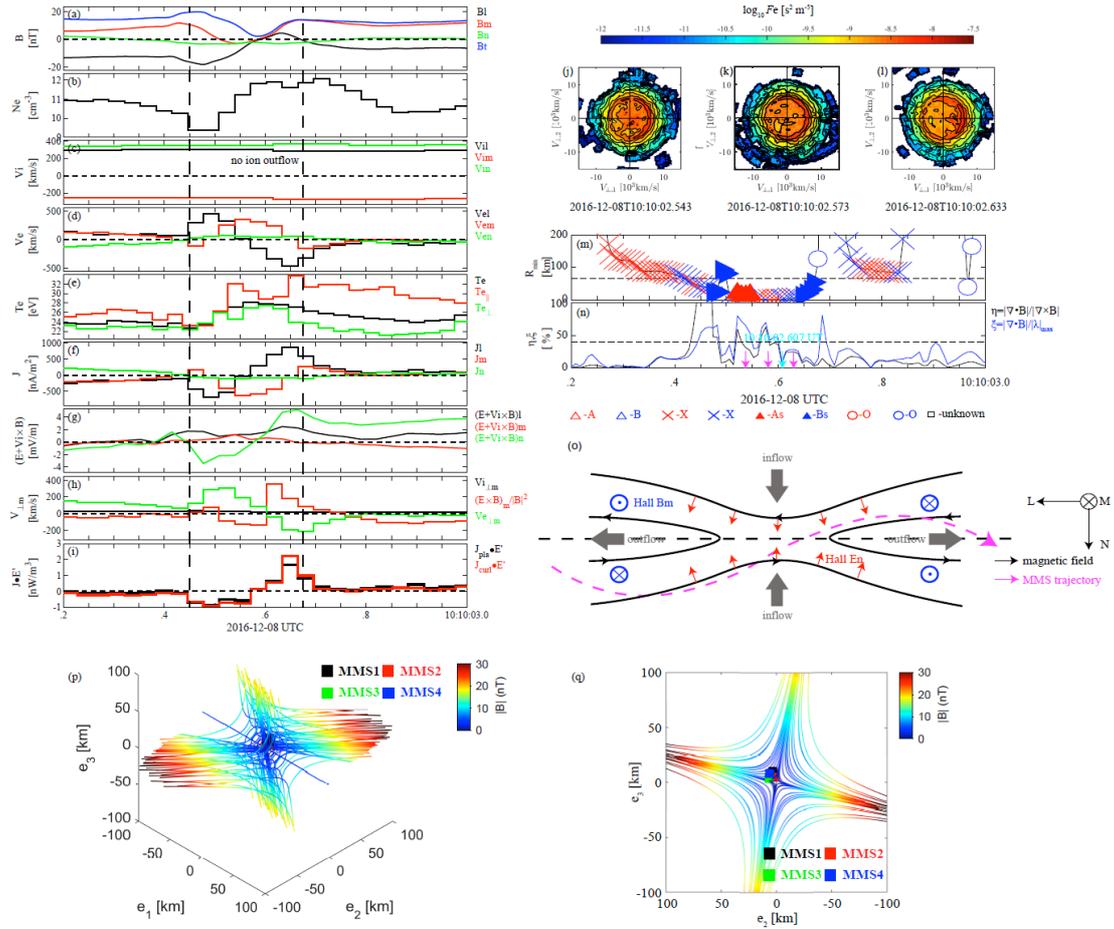

**Figure 3.** Details of case 1. (a) Magnetic field; (b) $N_e$; (c) $V_i$; (d) $V_e$; (e) $T_e$; (f) current density calculated by plasma moments; (g) $E + V_i \times B$; (h) M components of ion perpendicular velocity, $E \times B$ drift velocity, and electron perpendicular velocity; (i) $J \cdot E'$ (where $E' = E + V_e \times B$) calculated with plasma moments and curlometer method. The ion data and the electromagnetic field data used to calculate $J$ and $J \cdot E'$ are resampled to the electrons' resolution (i.e., 30 ms). All vectors in (a-i) are presented in LMN coordinates. (j-l) electron velocity distributions in the $V_{\perp 1}$-$V_{\perp 2}$ plane at different times, where $V_{\perp 1}$ is in the direction of $(b \times v) \times b$, and $V_{\perp 2}$ is along $b \times v$ ($b$ and $v$ are the measured unit vectors of the magnetic field and velocity moment, respectively); (m) $R_{min}$ with null types labeled, the horizontal black dashed line is the ion inertial length; (n) $\eta = |\nabla \cdot B|/|\nabla \times B|$ (black line) and $\xi = |\nabla \cdot B|/|\lambda|_{max}$ (blue line) to quantify the results of FOTE method, the horizontal black dashed line in (n) is 40%. The symbols at the bottom denote the 3D magnetic nulls (A, B, As, Bs) and the 2D approximation (X and O), the black square is the unknow null. The three magenta

arrows in (n) mark the times of electron velocity distributions, and the cyan arrow in (n) marks the time when the magnetic null's topology was reconstructed. (o) The inferred trajectory of MMS4 crossing the reconnection event; (p) 3D magnetic null's topology reconstructed by FOTE method and 2D view (q). Reconstruction is presented in the eigenvector coordinates ($e_1$, $e_2$, $e_3$). The eigenvectors in LMN coordinates are $e_1$=[0.10, 0.72, 0.69], $e_2$=[-0.11, 0.70, -0.71], and $e_3$=[-0.99, 0, -0.15]. The color scale denotes the strength of the magnetic field. The black, red, green, and blue rectangles represent four MMS spacecraft. Two vertical dashed lines mark the possible diffusion region.

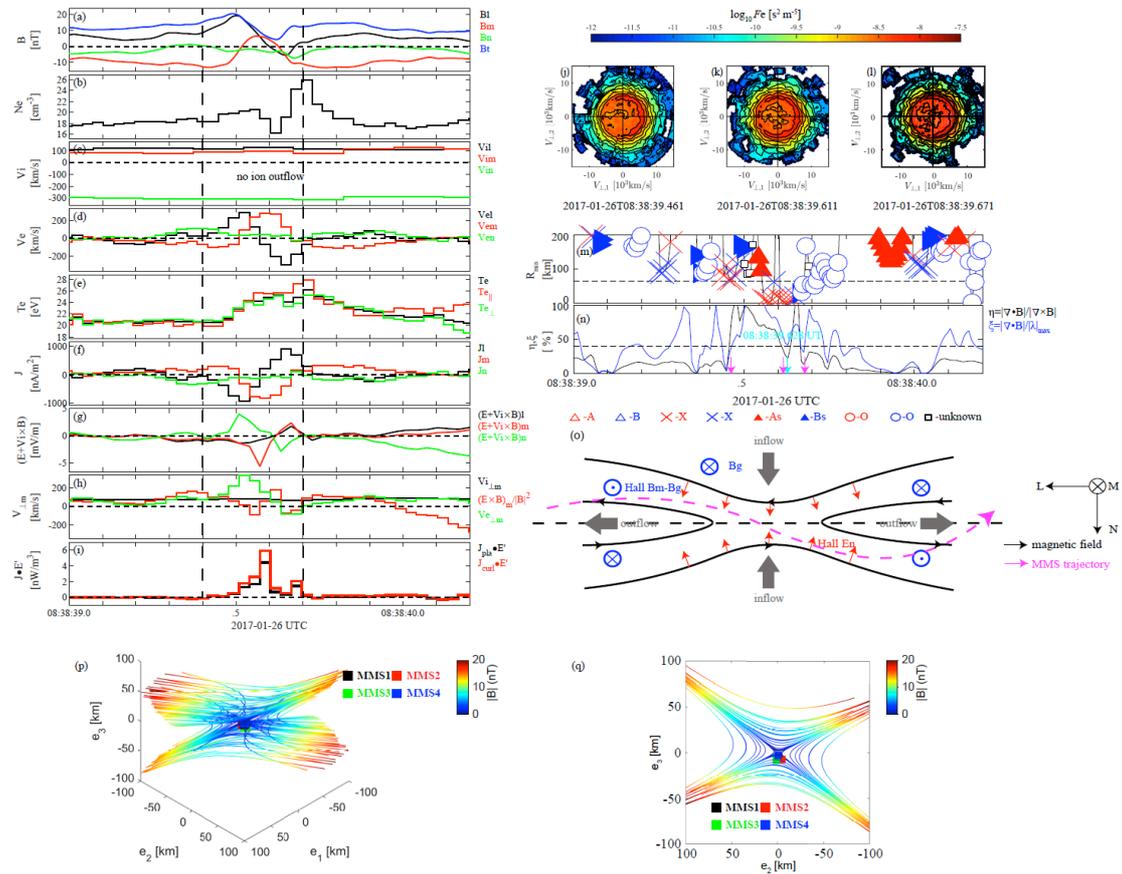

**Figure 4.** Details of case 2 with the same format of Figure 3. The eigenvectors in LMN coordinates are $e_1$=[0.66, -0.65, -0.37], $e_2$=[0.57, 0.76, -0.32], and $e_3$=[0.49, 0, 0.87].